\documentclass[12pt]{JHEP}
\usepackage{epsfig}

\def\appendix#1{
  \addtocounter{section}{1}
 \setcounter{equation}{0}
  \renewcommand{\thesection}{\Alph{section}}
 \section*{Appendix \thesection\protect\indent \parbox[t]{11.715cm} {#1}}
  \addcontentsline{toc}{section}{Appendix \thesection\ \ \ #1}
  }

\newcommand{\newsection}{
\setcounter{equation}{0}
\section}

\newcommand{\eq}[1]{\begin{equation} #1 \end{equation}}
\newcommand{\ar}[1]{\begin{eqnarray} #1 \end{eqnarray}}
\newcommand{\tr}{\mathop{\mathrm{tr}}\nolimits}

\def\e{{\,\rm e}\,}

\def\D{\delta}

\newcommand{\br}[1]{\left( #1 \right)}
\newcommand{\vev}[1]{\left\langle #1 \right\rangle}
\newcommand{\rf}[1]{(\ref{#1})}
\newcommand{\non}{\nonumber \\*}
\def\N{${\cal N}=4$ }
\def\l{\lambda}
\def\t{\sqrt{\lambda}}
\def\q{\theta}
\def\W{W(C_1,C_2)}
\def\f{\varphi}
\def\G{\Gamma}
\def\o{\omega}
\def\h{\frac{2\pi}{\sqrt{\lambda}}}

\preprint{ITEP--TH--9/01}

\title{String Breaking from Ladder Diagrams in SYM Theory}

\author{
K.~Zarembo\\
{\it Department of Physics and Astronomy}
\\{\it and Pacific Institute for the Mathematical Sciences}
\\{\it University of British Columbia}
\\ {\it 6224 Agricultural Road, Vancouver, B.C. Canada V6T 1Z1}
\\ \vskip .2 cm
and\\ \vskip .2cm
{\it Institute of Theoretical and Experimental Physics,}
\\ {\it B. Cheremushkinskaya 25, 117259 Moscow, Russia} \\ \vskip .5 cm
E-mail: {\tt zarembo@physics.ubc.ca}
}
\abstract{
The AdS/CFT correspondence establishes a string representation
for Wilson loops in \N SYM theory at large $N$ and large 't~Hooft coupling.
One of the clearest manifestations of the stringy behaviour
in Wilson loop correlators is the string breaking phase transition.
It is shown that resummation of planar diagrams without internal vertices
predicts the strong-coupling phase transtion in exactly the same
setting in which it arises from the string representation.
}

\begin{document}

\setcounter{page}{2}
\newsection{Introduction}

In its strongest form, the
 AdS/CFT correspondence establishes an equivalence  of
\N supersymmetric Yang-Mills (SYM) theory and string theory
in Anti-de-Sitter space \cite{Mal97}--\cite{Aha99}. Because
even a free string propagation in Anti-de-Sitter
space is rather complicated, going beyond the low-energy, supergravity
approximation in the AdS/CFT correspondence  is
extremely hard.
Not surprisingly,
the stringy nature of the AdS/CFT duality is not directly visible in
the supergravity limit. Fortunately, there is one exception:
Wilson loops in \N SYM
probe genuine stringy degrees of freedom even in the
supergravity regime \cite{Mal98,Rey98,Aha99,Dru99}.
Wilson loop correlators therefore should
display stringy
behavior in the large-$N$, large 't~Hooft coupling limit
of SYM theory which is dual to classical supergravity.

One of the clearest manifestations of the stringy behavior
in Wilson loop correlators is
the string breaking phenomenon. A good example of the string
breaking is Gross-Ooguri phase transition in the correlator
of two Wilson loops \cite{Gro98}--\cite{Kim01}.
When the loops are pulled apart, the string that connects them
eventually breaks and the correlation function
of the Wilson loops
undergoes a phase transition. This phase
transition looks rather unusual from the
field theory perspective: each Feynman diagram that contributes
to the Wilson loop correlator depends analytically on the
distance between the loops. Of course, intuition based on
individual Feynman graphs may well be wrong in the
 large 't~Hooft coupling limit.
To reach the strong coupling regime on the field theory side,
one has to sum all planar diagrams,
which is practically impossible in
an interacting field theory such as \N SYM.
It is thus rather surprising that partial resummation
that takes into account
only diagrams without internal vertices
gives results remarkably similar
to the supergravity predictions \cite{Eri00}--\cite{Ake01}.
For a circular Wilson loop,
diagrams without internal vertices reproduce
all available predictions of string theory,
including the area of classical string world-sheet \cite{Eri00'} and
the dimension of Teichm\"uller moduli space in string perturbation theory
\cite{Dru00}.
In fact,  the sum of diagrams without internal vertices
seems to give an exact result for the circular loop
to all orders of $1/N^2$ expansion
and for any 't~Hooft coupling due to special
conformal and supersymmetry transformation
properties of the circular loop operator \cite{Dru00}.

The diagrams without internal
vertices definitely do not exhaust all possible
contributions for other contours. For instance,
the large-$N$ expectation value for a pair of
anti-parallel Wilson lines receives contributions
from all possible planar diagrams (though there are some unexpected
cancellation in this case as well \cite{Eri00'}).  Nevertheless,
the sum of ladder diagrams for anti-parallel lines extrapolated
to the strong coupling limit
qualitatively agrees with the predictions of AdS/CFT duality
\cite{Eri00,Eri00'}. In particular,
the diagram resummation and the AdS/CFT correspondence predict the
same degree of screening of electric charge at large 't~Hooft coupling.
Similar results were found in non-commutative \N SYM theory \cite{Roz00}.

These observations support the conjecture
that  resummation of diagrams without internal vertices
always displays stringy behavior in the strong-coupling
regime.
To test this conjecture, I will sum up diagrams without internal vertices
for a correlator of two Wilson loops to see if this resummation gives rise
to the Gross-Ooguri phase transition.

The Gross-Ooguri phase transition in the correlator of Wilson
loops is reviewed in
Sec.~\ref{GO}. In sec.~\ref{ladder}, the same correlator is
analyzed in the ladder diagram approximation.

\newsection{Gross-Ooguri phase transition}\label{GO}

The Wilson loop operator that has right transformation
properties under supersymmetry \cite{Dru99}
couples not only to gauge potentials, but
also to the scalar fields, $\Phi_I$, $I=1\ldots 6$, of \N SYM theory
\cite{Mal98,Rey98}:
\eq{\label{php}
P(C)=\tr{\rm P}\exp\left[\oint_C d\tau\,\br{i A_\mu(x)\dot{x}_\mu
+\Phi_I(x)\theta_I|\dot{x}|}\right].
}
Here, $\theta_I$ is
a point on a five-dimensional unit sphere: $\theta^2=1$, and
$x_\mu(\tau)$ parameterizes contour $C$. The coupling to scalars
cancels potential UV divergences and Wilson loop correlators
are finite for smooth contours \cite{Dru99,Eri00'}.

The AdS dual of this operator is a world-sheet of
type IIB superstring that
propagates in the bulk of AdS space and whose ends
are attached to the contour $C$ on the boundary
\cite{Mal98,Rey98}. The tension of the
AdS
string is dimensionless and,
according to the AdS/CFT dictionary, is proportional to the square root of
the 't~Hooft coupling of  SYM theory:
$$T=\sqrt{\lambda}/2\pi,$$
$$\l=g^2_{\rm YM}N.$$
The large-$N$, large 't~Hooft coupling limit corresponds to a
free string with very large tension, which suppresses
all fluctuation of the string. Therefore, the
string world-sheet is classical in the strong coupling limit, and
Wilson loop correlators obey the minimal area law.

Actually, a straightforward implementation of the minimal area
law does not work because of the divergence of the area
due to a singular behavior
of the AdS metric at the boundary. It was argued that
the definition of the minimal area appropriate for
computation of Wilson loop correlators
involves the Legendre transform \cite{Dru99}.
An alternative regularization is
based on subtraction of the area of a reference surface with the same
boundary \cite{Mal98}.
In the semiclassical limit, these two regularizations are
mathematically equivalent. It
is not clear if this equivalence holds
beyond the semiclassical limit, probably it does not,
but at the semiclassical level, we are free to use either
of the two regularizations. The regularization by subtraction
then has an interesting consequence:
since the subtracted area is always larger than the
area of a minimal surface,  regularized area is always {\it
negative}. Hence,
\eq{\label{generalstring}
\ln\vev{P(C)}=\sqrt{\l}\times{\rm (positive~number)}
}
at large $\l$.
The AdS/CFT correspondence therefore predicts
that Wilson loop expectation values
 exponentiate at strong coupling and that the exponent is
proportional to $\t$ with positive coefficient.

The minimal surface is unique only for simplest contours. In general,
the area functional has several local minima, so the semiclassical
string partition function receives contributions from
several saddle points:
\eq{\label{ss}
\vev{P(C)}=\sum\alpha_i\exp\br{-\frac{\t\,A_i}{2\pi}},
}
where $A_i$ are (negative)
regularized areas of locally minimal surfaces and
$\alpha_i=\l^{-3/4}\times{\rm (power~series~in~}1/\t)$
represent quantum corrections due to fluctuations of the
string world sheet \cite{Dru00},\cite{Dru00'}--\cite{Kin00}.
At large $\l$, the term with the smallest area dominates.
Of course, each $A_i$ smoothly depends on geometric parameters
of the contour $C$, so the
Wilson loop expectation value is a smooth function of $C$, but
its large-$\l$ asymptotics, in general,  is not,
because different terms in the sum \rf{ss} may dominate
for contours of different shapes. If the shape of a contour is
continuously changed,
two local minima can become degenerate and
 the semiclassical partition function will
switch from one saddle point to another. The
large-$\l$ asymptotics of the Wilson loop will then
undergo a phase transition. This phenomenon is generic for
semiclassical amplitudes and
was encountered  for instance, in
the study of sphaleron transitions in quantum mechanics \cite{Chu92}
 or in quantum
field theory \cite{Hab96}--\cite{Bon00}.
In the context of string representation for Wilson loops,
the existence of a
phase transition due to rearrangement of minimal
surfaces was pointed out by Gross and Ooguri \cite{Gro98}, and has been
studied in much detail in \cite{Zar99,Ole00}.

\FIGURE{
\epsfxsize=7cm
\epsfbox{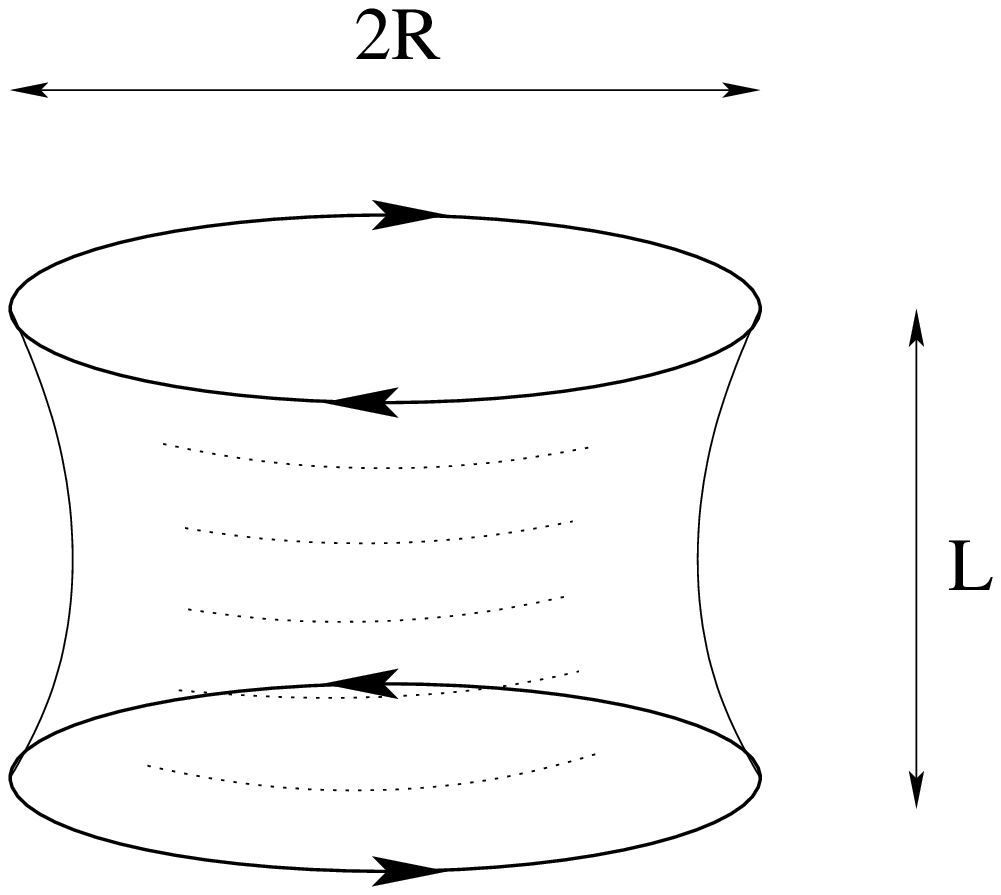}
\caption{Connected minimal surface.}
\label{ann}
}


The simplest correlation function that undergoes
the string-breaking phase transition is
the connected  correlator
of two Wilson loops:
\ar{\label{cc}
W(C_1,C_2)&=&\vev{P(C_1)P(C_2)}
\non &&-\vev{P(C_1)}\vev{P(C_2)},
}
where $C_1$ and $C_2$ are identical circles of opposite
orientation separated by distance $L$.
At strong coupling, the correlator is dominated by
the string world-sheet stretched between the two contours
(fig.~\ref{ann}). When the contours are pulled apart, the string will
eventually break, and the disconnected surface with the topology
of two disks (fig.~\ref{2d}) will become a global minimum. The
connectedness of the correlator is then achieved by perturbative exchange
of supergravity modes between separate minimal surfaces
\cite{Gro98,Ber98}.

\FIGURE{
\epsfxsize=7cm
\epsfbox{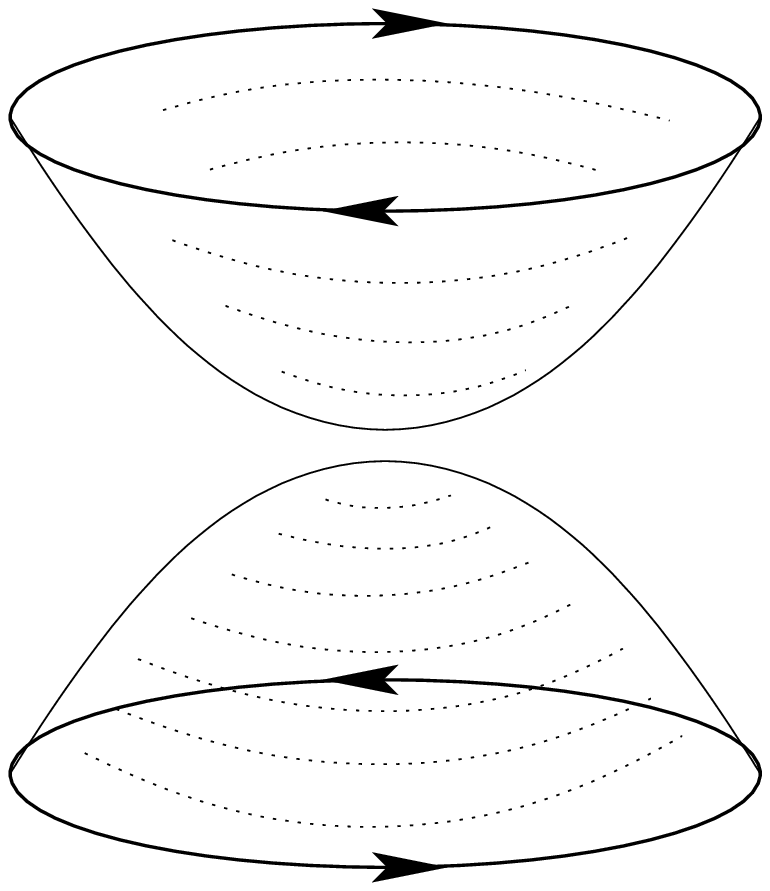}
\caption{Disconnected minimal surface.}
\label{2d}
}

The above intuitive arguments do not take into
account the strong curvature of $AdS_5\times S^5$ background in which
the strings propagate. However,
an explicit solution for semiclassical string world-sheet
\cite{Zar99}--\cite{Kim01}
shows that the curvature of AdS space does not alter
the qualitative picture of the Gross-Ooguri phase transition.
In fig.~\ref{are},
 the logarithm of the two-loop correlator (proportional to
minus an area of the minimal surface) computed in \cite{Zar99,Ole00}
is plotted as a function of the distance between the loops.

\FIGURE{
\epsfxsize=10cm
\epsfbox{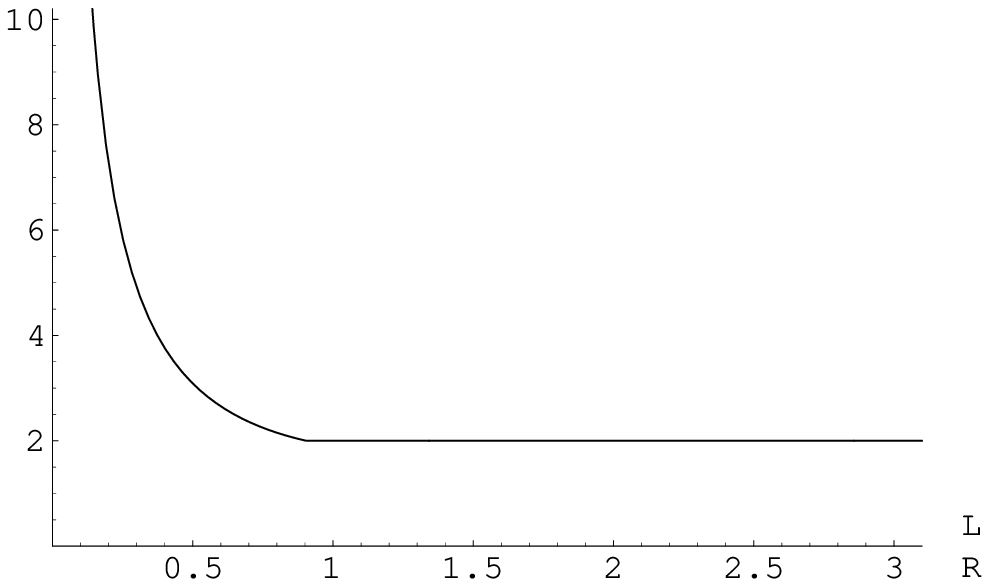}
\caption{$\ln W(C_1,C_2)/\sqrt{\l}$ as a function of the distance between
the loops.}
\label{are}
}

It is important to note that if the circles had the same orientation,
the connected minimal surface would not have existed. Consequently,
the phase transition takes place only for anti-parallel circles
and there is no phase transition when the circles are parallel.

\newsection{Ladder Diagrams}\label{ladder}

In this section, I will calculate the contribution of all planar
Feynman diagrams without internal vertices to the correlator of two
circular Wilson loops. This amounts in replacement
of the vacuum expectation value in
\rf{cc} by an average over free fields.
In the Feynman gauge, the Wick contraction of the exponent in the
Wilson loop operator is
\eq{\label{prp}
\vev{
\br{iA_\mu(x)\dot{x}_\mu
+\Phi_I(x)\theta_I|\dot{x}|}_{ij}
\br{i A_\mu(y)\dot{y}_\mu
+\Phi_I(y)\theta_I|\dot{y}|}_{kl}
}_0=\frac1N\,\D_{il}\D_{jk}\,\l\,
\frac{|\dot{x}| |\dot{y}|-\dot{x}\cdot\dot{y}}{8\pi^2|x-y|^2},
}
where $i,j,k,l$ are $U(N)$ group indices.

It is important to note that separation of all planar graphs in
the diagrams with and without
internal vertices is not gauge invariant and is
consistent only in the Feynman gauge, because only
in the Feynman gauge
these classes of diagrams are separately UV finite.
Any other gauge condition
brings in spurious divergences which invalidate
resummation of diagrams without internal vertices.
 The finiteness in the Feynman gauge is a consequence of
$SO(10)$ symmetry that rotates vector and scalar fields
of \N SYM and is inherited from
ten-dimensional Lorentz invariance. This symmetry
is broken by any other gauge condition, as well as
by interaction terms.

\FIGURE{
\epsfxsize=7cm
\epsfbox{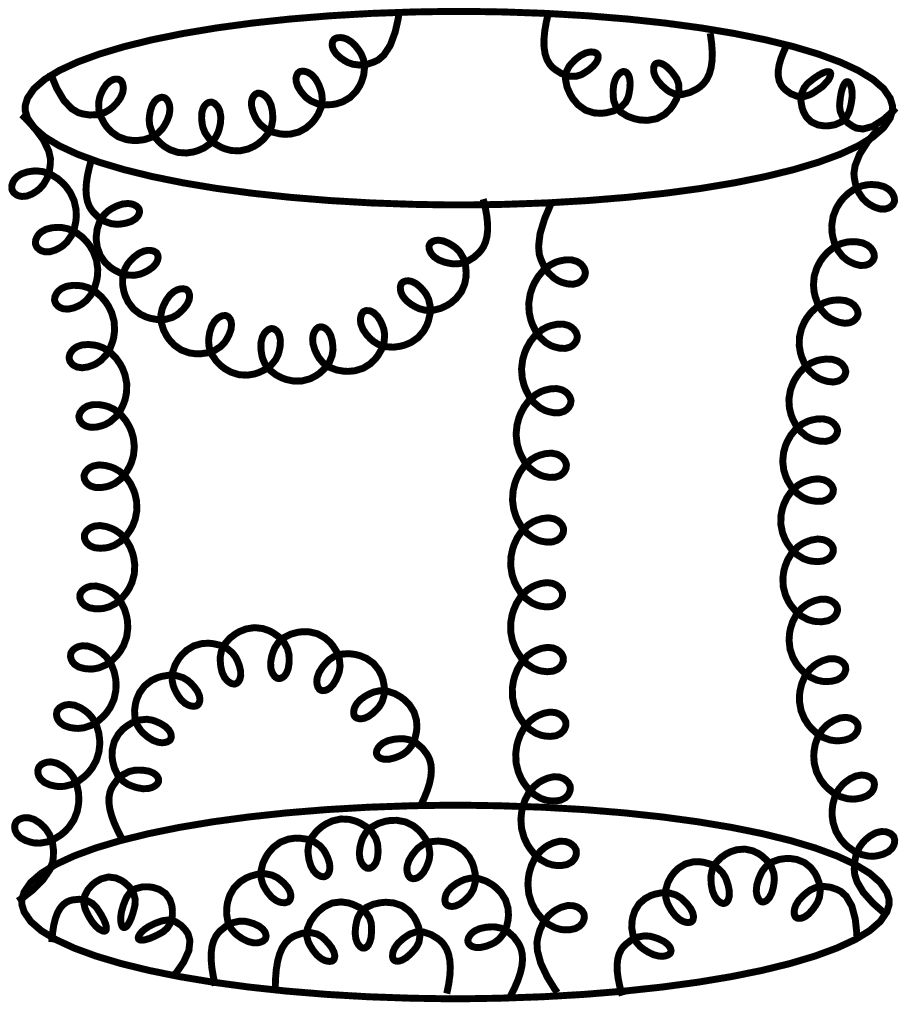}
\caption{A typical diagram that contributes to the connected
correlator of Wilson loops in the free-field approximation.}
\label{diagrams}
}

A typical planar diagram without internal vertices that
contributes to the connected correlator of two Wilson loops
(fig.~\ref{diagrams}) consists of
rainbow propagators, which are attached to one of the loops,
and ladder propagators, which connect the two loops together.
In spirit of the usual
identification of planar diagrams with discretized random surfaces
\cite{tHo74},
it is natural to associate ladder diagrams with connected string
world-sheets (fig.~\ref{ann}) and rainbow graphs with
the disconnected surfaces (fig.~\ref{2d}).
If the distance between
the circles is large compared to their radii, ladder diagrams
are suppressed. The large-$\l$ asymptotics is then governed by
exponentiation of rainbow graphs. The known sum of rainbow diagrams
\cite{Eri00'}
dictates the following asymptotic behavior
of the correlator at large separation between the contours:
\eq{\label{large}
W(C_1,C_2)\sim\vev{P(C_1)}\vev{P(C_2)}\approx\e^{2\t}~~~~~(L\gg R).
}
This exactly coincides with the AdS/CFT prediction
\cite{Ber98,Dru99},
fig.~\ref{are}. In particular, the
exponent in \rf{large} does not depend on $L$ or $R$.

In the opposite limit
of very small $L$ or of very large $R$,
the circles can be approximated by anti-parallel lines. In that
case, ladder diagrams evidently dominate.
The large-$\l$ extrapolation of their sum is also known
\cite{Eri00,Eri00'} and
implies the following asymptotics of the Wilson loop correlator
at small separation between the loops:
\eq{\label{short}
W(C_1,C_2)\sim\e^{2\t R/L }~~~~~(L\ll R).
}
The scaling with $\l$, $R$ and $L$ is again correct, but
the numerical coefficient in the exponent somewhat exceeds the
AdS/CFT prediction.

An exact resummation of ladder and rainbow diagrams
shows that asymptotics \rf{large} and \rf{short} do not match smoothly.
There is a phase transition at  $L_c=2 R$, at which
the correlator ceases to depend on the distance and the
asymptotics \rf{large} sets on. This phase transition can be associated with
breaking of the string made of ladder diagrams.
In the large-distance phase, the large-$\l$
behavior of the correlator is entirely determined by
rainbow graphs, while
in the short-distance phase, rainbow and ladder graphs are
equally important.

It is convenient to split the calculation into two parts:
first resum rainbow graphs, and then write down Dyson equation for
ladder diagrams. Let us denote the sum of the rainbow graphs
for an arc between polar angles $\q$ and $\q'$ by $W(\q'-\q)$:
\eq{
W(\q'-\q)=\raisebox{-1.5cm}{\epsfxsize=8cm\epsfbox{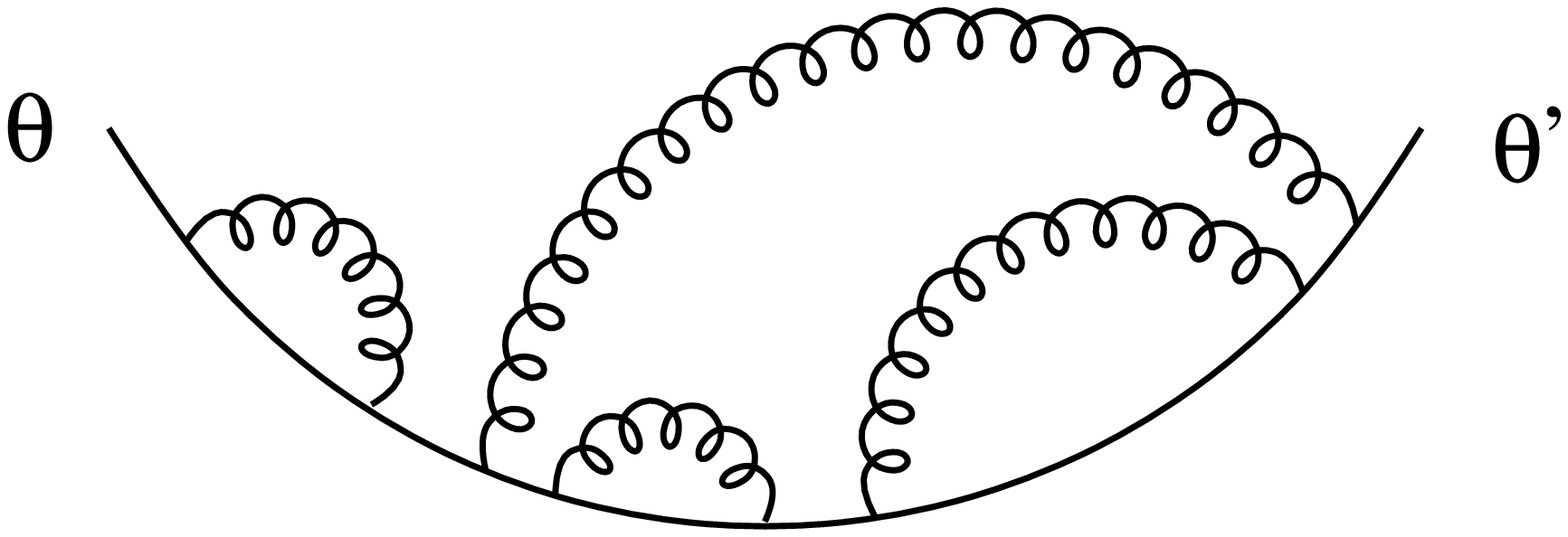}}.
}
Each rainbow propagator in this sum
is a constant, because,
 for a circular contour,
\eq{
\frac{|\dot{x}(\q_1)| |\dot{x}(\q_2)|
-\dot{x}(\q_1)\cdot\dot{x}(\q_2)}{|x(\q_1)-x(\q_2)|^2}
=\frac 12
}
independently of $\q_1$ and $\q_2$. The summation of
the rainbow graphs then
reduces to a zero-dimensional problem:
\eq{\label{mm}
W(s)=\vev{\e^{s M}},
}
where Gaussian
average over Hermitian $N\times N$ matrix $M$ is
defined to reproduce the SYM Wick contraction \rf{prp}:
\eq{
\vev{F(M)}=\frac 1Z\,\int dM\,F(M)\exp\br{-\frac{8\pi^2}{\l}\,N\tr M^2}.
}

The average \rf{mm} can be calculated using standard techniques of
large-$N$ random matrix models
\cite{Bre78}--\cite{Mak91}.
In fact, it is easier to find the Laplace transform
of $W(s)$:
\eq{\label{rnb}
W(z)\equiv\int_0^\infty ds\,\e^{-zs} \,W(s)=\vev{\frac{1}{z-M}}
=\frac{8\pi^2}{\l}\br{z-\sqrt{z^2-\frac{\l}{4\pi^2}}}.
}
The inverse Laplace transform yields:
\eq{
W(s)=\frac{4\pi}{\sqrt{\l}\, s}\,I_1\br{\frac{\sqrt{\l}\, s}{2\pi}},
}
where $I_1$ is the modified Bessel function.

For a ladder propagator that connects two loops,
\eq{\label{propag}
\frac{|\dot{x}_1(\q_1)| |\dot{x}_2(\q_2)|
-\dot{x}_1(\q_1)\cdot\dot{x}_2(\q_2)}{|x_1(\q_1)-x_2(\q_2)|^2}
=\frac12\,\frac{1+\cos\br{\q_1-\q_2}}
{\frac{L^2}{2R^2}+1-\cos\br{\q_1-\q_2}}\equiv G(\q_1-\q_2),
}
where $x_1(\q)$, $x_2(\q)$ parameterize the separate circles.
The connected correlator of Wilson loops contains at least one such
propagator, which we  extract
from the sum, for later convenience:
\eq{\label{wils}
\W=\frac{\l}{4\pi}\,\int_0^{2\pi} d\f\, G(\f)\G(2\pi,2\pi;\f).
}
All of the rest contributions can be
found by solving Dyson equation:
\eq{
\G(s,t;\f)=W(s)W(t)+\frac{\l}{8\pi^2}\,\int_0^s ds'\,\int_0^t dt'\,
W(s-s')W(t-t')G(s'-t'+\f)\G(s',t';\f),
}
supplemented by boundary conditions:
\eq{
\G(0,t;\f)=\G(s,0;\f)=0.
}
Iteration of this equation reproduces the sum of ladder diagrams with
all possible insertions of rainbow propagators.

Again, it is useful to do the Laplace transform:
\eq{
\G(z,w;\f)=\int_0^\infty ds\,\int_0^\infty dt\,\e^{-zs-wt}\,\G(s,t;\f),
}
after which the Dyson equation takes the form:
\eq{\label{dys}
\G(z,w;\f)=W(z)W(w)\br{1+\frac{\l}{8\pi^2}\sum_n \e^{in\f}G_n
\G(z-in,w+in;\f)
},}
where $G_n$ are Fourier modes of $G(\q)$:
\eq{
G(\q)=\sum_n G_n\e^{in\q}.
}
Singularities of the kernel
$\G(z,w;\f)$ at complex $z$ and $w$ essentially
determine its inverse Laplace transform. To get an
idea of the range of $z$ and $w$ in which the singularities
occur, let us consider
an iterative solution of the Dyson equation \rf{dys}. To the first
approximation, the kernel
factorizes on two separate
sums of rainbow diagrams: $\G(z,w;\f)=W(z)W(w)$.
 The singularities are branch cuts across
the real axes in the complex $z$ and $w$ planes with branch points
at $\pm\t/2\pi$. A first iteration of eq.~\rf{dys} will shift cuts
into the complex plane and branch points at $\pm\t/2\pi+in$ will arise
 for any integer $n$.
Next iterations do not produce any new singularities. In the most
interesting regime of large $\l$, the branch cuts extend to large
distances of order of $\t$ along the real axis. It is therefore
convenient to rescale $z$ and $w$ by $\t/2\pi$. The form of
eq.~\rf{dys} then suggests the following change of variables:
\eq{\label{defl}
\G\br{\frac{\t}{2\pi}(\o+ip),\frac{\t}{2\pi}(\o-ip);\f}=
\frac{4\pi^2}{\l}\,\e^{\frac{i\t p\f}{2\pi}}L(\o,p;\f). }
Introducing the notation: \eq{ D(\o)\equiv
\frac{2\pi}{\t}\,\frac{1}{W\br{\frac{\t}{2\pi}\,\o}} =\frac
12\br{\o+\sqrt{\o^2-1}}, } we can rewrite the Dyson equation as \eq{
D(\o+ip)D(\o-ip)L(\o,p;\f) -\frac 12\sum_n G_n L\br{\o,p-\frac{2\pi
n}{\t};\f}= \e^{-i\frac{\sqrt{\lambda }\,p\f }{2\pi }}. } The
Fourier transform in $p$, \eq{\label{four}
L(\o,x;\f)=\frac{\t}{4\pi^2}\int_{-\infty}^{+\infty}
dp\,\e^{\frac{i\t px}{2\pi}} L(\o,p;\f), } then yields a
Schr\"odinger-like equation: \eq{\label{sch}
\left[D\br{\o+\h\,\frac{d}{dx}}D\br{\o-\h\,\frac{d}{dx}}-\frac
12\,G(x) \right]L(\o,x;\f)=\D(x-\f) } with the Hamiltonian
\eq{\label{ham} H(p,x;\o)= D(\o+ip)D(\o-ip)-\frac 12 G(x), } where
the momentum operator is defined as \eq{ p=-i\,\h\,\frac{d}{dx}. }

The formal solution of the Schr\"odinger equation \rf{sch} in the
coordinate representation is \eq{ L(\o,x;\f)=\vev {x | H^{-1}(\o)
|\f}. }
In terms of the complete set of eigenfunctions of $H$\footnote{Since
the potential $G(x)$ is periodic, the spectrum of $H$ forms a band
structure and the summation over eigenvalues, strictly speaking,
should be understood as the integration weighted with the density of
states.}: \eq{\label{eigenfunc}
L(\o,x;\f)=\sum_n\frac{\psi^*_n(\f;\o)\psi_n(x;\o)}{E_n(\o)}\,. }
The wave functions $\psi_n$ are properly normalized solutions of the
Schr\"odinger equation: \eq{\label{sse}
H\br{-i\h\,\frac{d}{dx},x;\o}\psi_n(x;\o)=E_n(\o)\psi_n(x;\o). }

The kernel $\G$ with coinciding arguments determines the expectation
value of the Wilson loop correlator, according to \rf{wils}. Its
Laplace transform can be easily found with the help of \rf{defl},
\rf{four}: \ar{ \int_0^\infty ds\,\e^{-\frac{\t \o s}{\pi}}
\G(s,s;\f) &=&\frac{\t}{4\pi^2}\int_{-\infty}^{+\infty} dp\,
\G\br{\frac{\t}{2\pi}(\o-ip),\frac{\t}{2\pi}(\o+ip);\f} \non
&=&\frac{4\pi^2}{\l}\,L(\o,\f;\f). } The inverse Laplace transform
gives: \eq{\label{gss}
\G(s,s;\f)=\frac{2}{i\t}\int_{C-i\infty}^{C+i\infty}
d\o\,\e^{\frac{\t\o s}{\pi}}L(\o,\f;\f), } where the contour of
integration passes all singularities of the integrand on the right.
Substitution of the solution of the Schr\"odinger equation
\rf{eigenfunc} into \rf{gss} gives for the Wilson loop expectation
value: \eq{\label{main} \W=\frac{\t}{2\pi
i}\int_{C-i\infty}^{C+i\infty} d\o\, \e^{2\t\o}\sum_n
\frac{1}{E_n(\o)}\int_0^{2\pi}d\f\, G(\f)\left|\psi_n(\f;\o)
\right|^2. }

This expression is valid for any $\l$ and, in particular, for large
$\l$. It actually simplifies in the strong-coupling limit, because
then the integrand in \rf{main} rapidly oscillates and the integral
over $\o$ is saturated by the singularity of the integrand in the
complex $\o$ plane with the largest real part\footnote{It can be
shown that such singularity lies on the positive real semi-axis.}:
\eq{\label{exp} \W\simeq\e^{2\t\o_0}. } Therefore, diagrams without
internal vertices exponentiate, and the exponent is proportional to
$\t$ with positive coefficient, in agreement with general properties
of the AdS/CFT prediction.

It is not hard to find $\o_0$ in the limit of large $\l$.
There are two sources of non-analyticity in the integrand
of \rf{main}: (i) when $E_n(\o)$
hits zero, the integrand  develops a pole\footnote{To be
more precise, the energy spectrum is continuous, so the poles
associated with distinct energy levels fuse and form a cut.}
and (ii) each $E_n(\o)$ is a multivalued function
because the Hamiltonian \rf{ham} is not analytic in $\o$.

The reason for simplification at large $\l$ stems from
the commutation relation
\eq{
[x,p]=i\h,
}
which
shows that  $2\pi/\t$ plays the role of the Plank constant.
The large-$\l$ limit is therefore semiclassical. The semiclassical spectrum
of the Hamiltonian \rf{ham}, which is self-adjoined
at real $\o$, forms a continuum
that starts from the minimum of
the classical energy in the phase space:
\eq{
\lim_{\l\rightarrow\infty}E_0(\o)=\min_{p,x} H(p,x;\o).
}
The kinetic energy is minimal
at zero momentum. The minimum of the potential
is reached at
$x=0$, so
\eq{
E_0(\o)\approx \Bigl(D(\o)\Bigr)^2-\frac12\,G(0)=
\frac 14\,\br{\o+\sqrt{\o^2-1}}^2-\frac{R^2}{L^2}.
}

The ground state energy always has a branch point at $\o_0=1$.
This singularity originates from the square root branch point
in the sum of rainbow graphs \rf{rnb} at $z=\t/2\pi$ and
translates into the distance-independent large-$\l$
asymptotics \rf{large} for the Wilson loop correlator.
Another singularity
arises when $E_0$
crosses zero. This happens at
$$
\o_0=\frac{R}{L}+\frac{L}{4R},
$$
provided that  $L<2R$, otherwise
eigenvalues of the Hamiltonian \rf{ham} are
positive for any $\o>1$. 
At distances larger than $L_c=2R$, the branch point at
$\o=1$ is the only singularity of the integrand in
\rf{main}. At smaller distances,
the integrand has a pole at larger $\o$ in addition to the branch cut.
Thus
\eq{
\o_0=\left\{
\begin{array}{ll}
1,&~L>2R\\
\frac{R}{L}+\frac{L}{4R},&~L<2R
\end{array}
\right.\,,
}
and, consequently,
\eq{
\W\simeq\left\{
\begin{array}{ll}
\e^{2\t},&~L>2R\\
\e^{\t\br{\frac{2R}{L}+\frac{L}{2R}}},&~L<2R
\end{array}
\right.\,.
}

\FIGURE{
\epsfxsize=10cm
\epsfbox{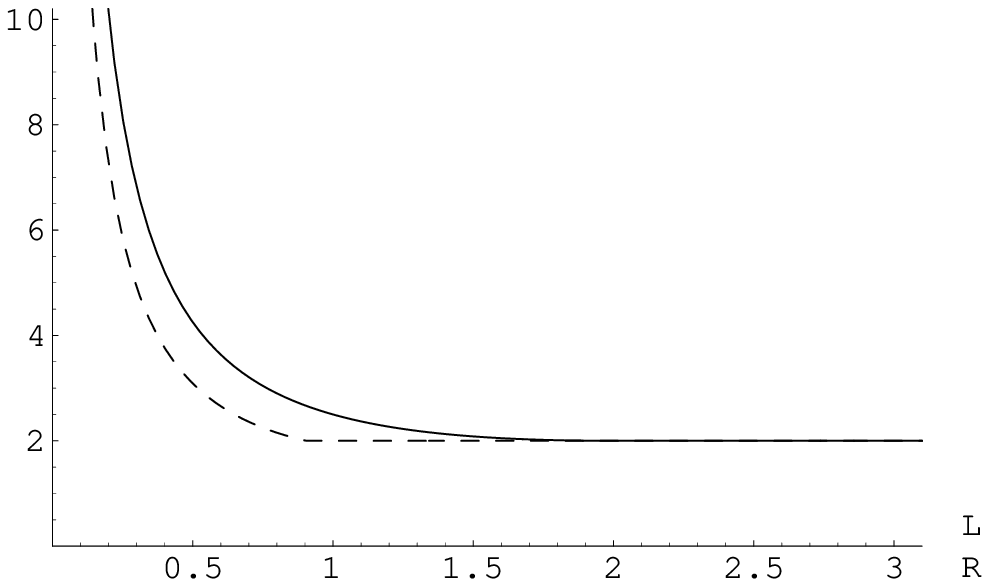}
\caption{$\ln W(C_1,C_2)/\sqrt{\l}$ as a function of the distance between
the loops. The solid curve represents the result of resummation
of diagrams without internal vertices extrapolated to the strong
coupling. The dashed curve is the AdS/CFT prediction.}
\label{are+}
}

Thus, the Wilson loop expectation value undergoes a phase transition at
large $\l$ in the ladder diagram
approximation. This phase transition is completely analogous to the
Gross-Ooguri transition in the semiclassical string amplitude.
At large distances, the expectation value is dominated by
rainbow diagrams that can be associated with disconnected
string world-sheets. The field theory calculation agrees exactly with
the prediction of AdS/CFT correspondence in this case for the reasons
explained in \cite{Dru00}. At short distances,
ladder graphs, which are counterparts of the connected world-sheets,
become increasingly important. Despite
the lack of apparent reasons
for the field theory calculation to be accurate
in the short-distance
phase, the result of the diagram resummation
only slightly deviates from the
AdS/CFT prediction (fig.~\ref{are+}).

It is straightforward to repeat resummation
of the diagrams without
internal vertices for parallel circles.
The inversion of the orientation changes sign in
the numerator of \rf{propag}:
\eq{\label{propp}
\tilde{G}(\q_1-\q_2)\equiv\frac{|\dot{x}_1(\q_1)| |\dot{x}_2(\q_2)|
-\dot{x}_1(\q_1)\cdot\dot{x}_2(\q_2)}{|x_1(\q_1)-x_2(\q_2)|^2}
=\frac12\,\frac{1-\cos\br{\q_1-\q_2}}
{\frac{L^2}{2R^2}+1-\cos\br{\q_1-\q_2}}.
}
All subsequent calculations remain the same up to
replacement of $G$ by $\tilde{G}$. In particular, \rf{main}
still holds, but
the
large-$\l$ limit of the ground state energy now is
given by
\eq{
\tilde{E}_0(\o)\approx \Bigl(D(\o)\Bigr)^2-\frac12\,\tilde{G}(\pi)=
\frac 14\,\br{\o+\sqrt{\o^2-1}}^2-\frac{R^2}{L^2+2R^2},
}
because  the potential, $-\tilde{G}(x)$,
has a minimum at $x=\pi$.
This expression turns out to be positive for any $R$ and $L$,
which means that energy levels never cross zero, so the only source
of non-analyticity in $\o$ is the branch point associated with
rainbow diagrams. Consequently, there is no
string-breaking phase transition, in agreement with what
is expected from AdS/CFT correspondence.
In fact, the string theory prediction for parallel circles
is reproduced exactly, since
rainbow graphs always dominate:
\eq{
W(C_1,\tilde{C}_2)\approx\e^{2\t}.
}

\newsection{Discussion}

Retaining only Feynman graphs without internal vertices
is well motivated at strong coupling only in a special case of the circular
Wilson loop. However, resummation of such diagrams bears a qualitative
agreement with AdS/CFT correspondence for all Wilson loop
correlators studied so far.
The strong coupling
asymptotics of the resummed perturbative series is always of the
form \rf{generalstring}, which is a general prediction of the string theory.
In the case of the two-loop correlator,
the resummed ladder diagrams undergo a strong-coupling phase transition
when we expect the string breaking
to occur and depend analytically
on the distance between the loops when  string breaking does not happen.
Results of perturbative calculation
do not deviate much from the AdS/CFT prediction
even quantitatively, which
allows us to speculate that resummation of
diagrams without
internal vertices
may in general constitute a first approximation of some systematic expansion,
and that
there may be a more direct link
between planar diagrams without internal vertices and strings.

\acknowledgments

I would like to thank G.~Semenoff and C.~Thorn for discussions and
N.~Obers for useful correspondence. This work was supported by NSERC
of Canada, by Pacific Institute for the Mathematical Sciences and in
part by RFBR
 grant 98-01-00327 and RFBR grant
00-15-96557 for the promotion of scientific schools.


\end{document}